\documentclass[3p,times,procedia]{elsarticle}
\usepackage{nupha_ecrc}

\volume{00}

\firstpage{1}

\journalname{Nuclear Physics A}

\runauth{W. Florkowski et al.}

\jid{nupha}

\jnltitlelogo{Nuclear Physics A}

\usepackage{amssymb}
\biboptions{comma,square,sort&compress}

\usepackage[figuresright]{rotating}

\usepackage{nicefrac}
\usepackage{slashed}
\usepackage{amsmath,graphicx}
\usepackage[colorlinks=true,linktocpage=true,linkcolor=blue,citecolor=blue]{hyperref}
\usepackage{float}
\usepackage{nicefrac}
\usepackage[normalem]{ulem}
\usepackage{amsmath}
\usepackage{subcaption}

\def\k{{\boldsymbol k}}

\def\dd{{\rm d}}

\def\trel{\tau_{\rm rel}}

\def\vsound{c_s}

\def\azeta1{a^{(n)}_{\zeta_1}}

\def\s2{\sqrt{2}\,}

\newcommand{\onethird}{{\nicefrac{1}{3}}}

\newcommand{\smallG}{{\rm \scriptscriptstyle{G}}}

\newcommand{\GZ}{{\rm \scriptscriptstyle{GZ}}}

\newcommand{\BE}{{\rm \scriptscriptstyle{BE}}}
\newcommand{\beq}{\begin{eqnarray}}
\newcommand{\eeq}{\end{eqnarray}}
\newcommand{\be}{\begin{eqnarray*}}
\newcommand{\ee}{\end{eqnarray*}}
\newcommand{\bqa}{\begin{eqnarray}}
\newcommand{\eqa}{\end{eqnarray}}

\newcommand{\K}{{\bf k}}

\begin{document}
%%%%%%%%%%%%%%%%%%%%%%%%%%%%%%%%%%%%%%%%%%%%%%%%%%%%%%%%%%%%%%%%%%%%%%%%%%
%%%%%%%%%%%%%%%%%%%%%%%%%%%%%%%%%%%%%%%%%%%%%%%%%%%%%%%%%%%%%%%%%%%%%%%%%%
\begin{frontmatter}
%%%%%%%%%%%%%%%%%%%%%%%%%%%%%%%%%%%%%%%%%%%%%%%%%%%%%%%%%%%%%%%%%%%%%%%%%%

\dochead{}

\title{Strong-coupling effects in a plasma of confining gluons}

\author[a]{Wojciech Florkowski}
\ead{wojciech.florkowski@ifj.edu.pl}
\author[a]{Radoslaw Ryblewski}
\ead{radoslaw.ryblewski@ifj.edu.pl}
\author[b]{Nan Su}
\ead{nansu@th.physik.uni-frankfurt.de}
\author[c]{Konrad Tywoniuk}
\ead{konrad@ecm.ub.edu}

\address[a]{The H. Niewodnicza\'nski Institute of Nuclear Physics, Polish Academy of Sciences, PL-31342 Krak\'ow, Poland}
\address[b]{Institut f\"ur Theoretische Physik, Goethe-Universit\"at Frankfurt, D-60438 Frankfurt am Main, Germany}
\address[c]{Deptartamento d'Estructura i Constituents de la Mat\`eria, Institut de Ci\`encies del Cosmos (ICCUB), Universitat de Barcelona, Mart\`i Franqu\`es 1, E-08028 Barcelona, Spain}

\begin{abstract}
The plasma consisting of confining gluons resulting from the Gribov  quantization of the SU(3) Yang-Mills  theory is studied using non-equilibrium fluid dynamical framework. Exploiting the Bjorken symmetry and using linear response theory a general analytic expressions for the bulk, $\zeta$, and shear, $\eta$, viscosity coefficients are derived. It is found that the considered system exhibits a number of properties similar to the strongly-coupled theories, where the conformality is explicitly broken. In particular, it is shown that, in the large temperature limit, $\zeta/\eta$ ratio, scales linearly with the difference $\onethird - c_s^2$, where $c_s$ is the speed of sound. Results obtained from the analysis are in line with the interpretation of the quark-gluon plasma as an almost perfect fluid.
\end{abstract}


%%%%%%%%%%%%%%%%%%%%%%%%%%%%%%%%%%%%%%%%%%%%%%%%%%%%%%%%%%%%%%%%%%%%%%%%%%
\end{frontmatter}
%%%%%%%%%%%%%%%%%%%%%%%%%%%%%%%%%%%%%%%%%%%%%%%%%%%%%%%%%%%%%%%%%%%%%%%%%%

%%%%%%%%%%%%%%%%%%%%%%%%%%%%%%%%%%%%%%%%%%%%%%%%%%%%%%%%%%%%%%%%%%%%%%%%%%
\section{Introduction}
\label{s-int}
%%%%%%%%%%%%%%%%%%%%%%%%%%%%%%%%%%%%%%%%%%%%%%%%%%%%%%%%%%%%%%%%%%%%%%%%%%
%
\par The enormous amount of data collected in the ultra-relativistic heavy-ion collision experiments in the last decade strongly suggests that a new strongly-coupled state of nuclear matter, the so called Quark-Gluon Plasma (QGP), is produced. The QGP's space-time evolution turns out to be very well described within the effective framework of relativistic fluid dynamics, with the properties of the system encoded in the equation of state and transport coefficients. These properties should, in principle, follow from the underlying theory of Quantum Chromodynamics (QCD), using, for example, numerical QCD  simulations on the lattice  (lQCD) \cite{Borsanyi:2013bia,Asakawa:2013laa,Bazavov:2014pvz} or Hard-Thermal-Loop resummed perturbation theory \cite{Andersen:2011sf,Mogliacci:2013mca,Haque:2014rua,Su:2015esa}. However, while the lQCD results on thermodynamic variables are quite precise nowadays, the results concerning the transport coefficients, in particular bulk, $\zeta$, and shear, $\eta$, viscosities, are still plagued with uncertainties which are too large to make firm statements \cite{Meyer:2007dy,Meyer:2007ic}. In view of these arguments any hints about transport properties of the QGP are more than welcome and  have to be drawn from other theoretical considerations.
\par In this proceedings contribution we shortly review our main results of Refs.~\cite{Florkowski:2015rua,Florkowski:2015dmm}, where, for the first time, the in- and out-of-equilibrium properties of a plasma consisting of confining gluons were studied. The plasma constituents were described by the dispersion relation resulting from the Gribov-Zwanziger (GZ) quantization procedure \cite{Gribov:1977wm,Zwanziger:1989mf} of the SU(3) Yang-Mills (YM)  theory. In this way we attempt to study implications of non-Abelian structure of the gluon system on the dynamic properties of the plasma. In particular we focus on the bulk and shear viscosity coefficients, which are crucial for application of the viscous fluid dynamical modelling. 
%
%%%%%%%%%%%%%%%%%%%%%%%%%%%%%%%%%%%%%%%%%%%%%%%%%%%%%%%%%%%%%%%%%%%%%%%%%%
\section{The Gribov dispersion relation}
\label{s-gz}
%%%%%%%%%%%%%%%%%%%%%%%%%%%%%%%%%%%%%%%%%%%%%%%%%%%%%%%%%%%%%%%%%%%%%%%%%%
%
The GZ approach is based on a meticulous gauge fixing procedure  of the quantized YM theory introduced by Gribov in Ref.~\cite{Gribov:1977wm}, which results in the introduction of a new scale, the so called Gribov parameter, $\gamma_\smallG$, in the framework. The $\gamma_\smallG$ parameter controls the onset of the confinement effects in the gluon system, and it is found from the self-consistent solution of the gap-equation \cite{Gribov:1977wm,Zwanziger:1989mf}. As a result the dispersion relation of interacting gluons, which in the Coulomb gauge reads \cite{Gribov:1977wm}
\begin{equation}\label{disp}
E(\K) = \sqrt{\K^2 + \frac{\gamma_\smallG^4}{\K^2}},
\end{equation}
improves the behavior of the theory in the infrared regime.  
The theory was generalized to finite temperatures by Zwanziger in Ref.~\cite{Zwanziger:2004np}, where its qualitative agreement with the pure glue lattice results was shown. Recently, the GZ approach received significant interest from the theory side resulting in several interesting findings of QCD under extreme conditions \cite{Zwanziger:2006sc,Kojo:2009ha,Kojo:2012js,
Fukushima:2013xsa,Su:2014rma,Kharzeev:2015xsa,Canfora:2015yia,
Bandyopadhyay:2015wua}, which make this approach a promising playground for the study of confinement effects in YM realtime dynamics.
%%%%%%%%%%%%%%%%%%%%%%%%%%%%%%%%%%%%%%%%%%%%%%%%%%%%%%%%%%%%%%%%%%%%%%%%%%
\section{Bulk and shear viscosity of confining gluons}
\label{s-res}
%%%%%%%%%%%%%%%%%%%%%%%%%%%%%%%%%%%%%%%%%%%%%%%%%%%%%%%%%%%%%%%%%%%%%%%%%%
%
\par In order to apply our fluid dynamical considerations we first rewrite Eq.~(\ref{disp}) in the Lorentz covariant form, $E(k \cdot u) = \sqrt{(k \cdot u)^2 + \gamma_\smallG^4/(k \cdot u)^2 }$, where $\k$ is the three-momentum of a gluon, $k^0=|\k|$, and $u^{\mu}$ is the four-velocity of the rest frame. The latter formula may then be used in the finite temperature results for thermodynamic variables found by Zwanziger in Ref.~\cite{Zwanziger:2004np}. For the calculation of  transport coefficients, without loss of generality,  we may impose  the Bjorken symmetry \cite{Bjorken:1982qr} of the system, that is we consider simple longitudinally boost-invariant and transversally homogeneous (0+1)-dimensional expansion. In this case $u^{\mu}=x^{\mu}/\tau$ and all thermodynamic variables become functions of proper time, $\tau =\sqrt{t^2-z^2}$, solely.
\par Our approach is based on the relaxation time approximation (RTA) for the collision kernels \cite{Florkowski:2013lya}. Considering small perturbations around the equilibrium distribution function, $f = f_\GZ + \delta f$, and making use of the first-order viscous fluid dynamical expressions for bulk and shear viscous pressure corrections, after straightforward algebra, we find the following expressions  \cite{Florkowski:2015rua,Florkowski:2015dmm}
\begin{equation}
\label{eq:ZetaSimplified}
\zeta = \frac{g_0\gamma_\smallG^5 }{3\pi^2} \frac{\tau_{\rm rel}}{T}  \int_0^\infty \dd y \,\left[ \vsound^2 - \frac{1}{3}\frac{y^4 - 1}{y^4+ 1} \right] f_\GZ(1+f_\GZ),
\end{equation}
\begin{equation}
\label{eq:EtaSimplified}
\eta =\frac{g_0\gamma_\smallG^5}{30\pi^2} \frac{\trel}{T} \int_0^\infty \dd y \,\frac{\left( y^4-1 \right)^2}{y^4+1} f_\GZ(1+f_\GZ) \,,
\end{equation}
for the bulk and shear viscosity, respectively. Above $f_\GZ = \{\exp[\gamma_\smallG\sqrt{y^2+y^{-2}}/T]-1 \}^{-1}$, $c_s$ is the speed of sound, $g_0$ is the degeneracy factor and $\trel$ is the relaxation time.
\par
\begin{figure}[t]
\begin{center}
\includegraphics[scale=0.49]{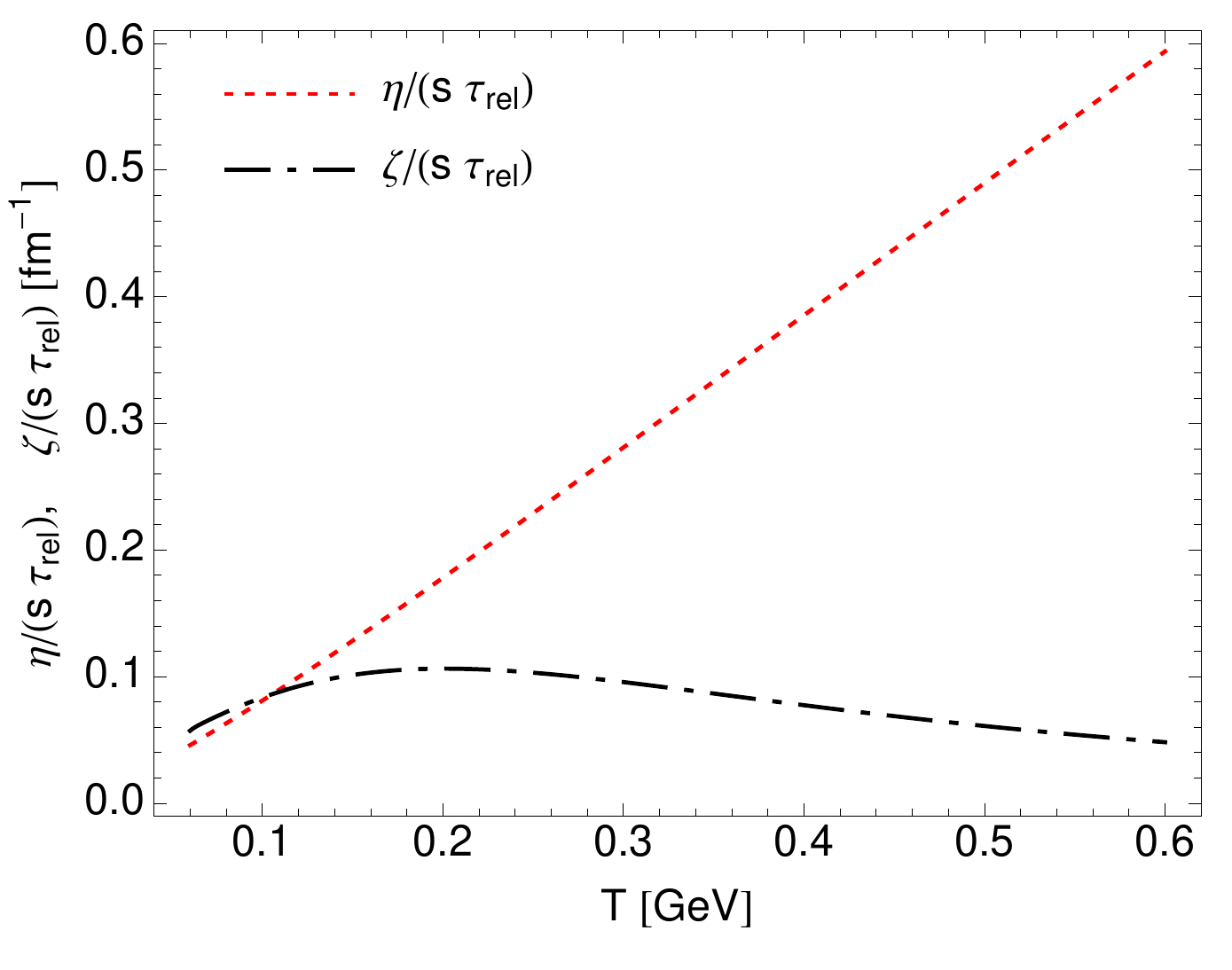}
\includegraphics[scale=0.49]{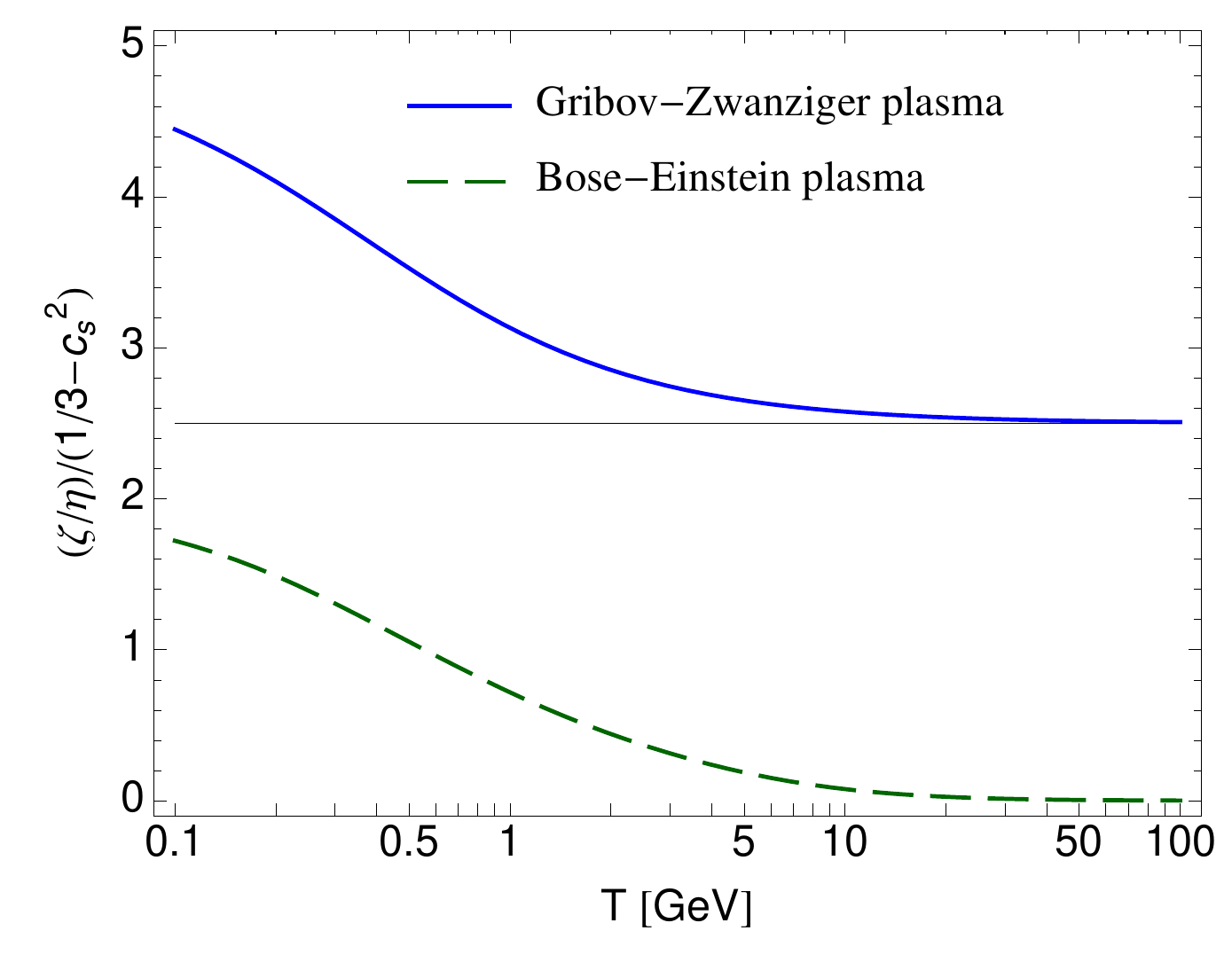}
\end{center}
\caption{Left panel: The temperature dependence of bulk (black dashed-dotted line) and shear (red dotted line) viscosity obtained within the linear response theory, Eqs.~(\ref{eq:ZetaSimplified})--(\ref{eq:EtaSimplified}), scaled by the equilibrium entropy density and the relaxation time. Right panel: The $\zeta/\eta / \left(1/3 -c_s^2\right)$ scaling as a function of temperature for the Gribov-Zwanziger plasma (solid blue line) and the massive Bose-Einstein plasma (dashed green line).
}%
\label{fig:trans}
\end{figure}

 In the left panel of Fig.~\ref{fig:trans} we present the temperature dependence of $\zeta$ (dashed-dotted line) and $\eta$ (dotted line) calculated from Eqs.~(\ref{eq:ZetaSimplified})-(\ref{eq:EtaSimplified}) and scaled by the equilibrium entropy density, $s$, and the relaxation time. Similarly to the ideal massless gas one observes approximately linear scaling of scaled shear viscosity, \mbox{$\eta/s \sim T \trel$}, with temperature. Another interesting observation is the enhancement of scaled bulk viscosity around the critical temperature $T_c=260$ MeV, which highlights the importance of bulk viscosity for the heavy-ion phenomenology due to cavitation in the plasma \cite{Sanches:2015vra}. Finally, unlike in the quasiparticle models, at low temperatures the bulk and shear viscosity become comparable.
\par In the right panel of Fig.~\ref{fig:trans} we plot the ratio $\zeta/\eta$ scaled by the factor $1/3 -c_s^2$ vs. temperature for the GZ plasma (solid line) and a massive non-interacting Bose-Einstein (BE) plasma (dashed line). We note that  the ratio $\zeta/\eta$ is independent of the relaxation time. We observe that the values predicted within GZ formalism are always larger than the ones obtained for BE plasma. Moreover, we see that the scaling is approximately linear at asymptotically large temperatures. 
\par The latter can be shown analytically by performing large-$T$ expansions of the shear and bulk viscosities. In this way for  $T \gg \gamma_\smallG$ we arrive at the following scaling
\begin{equation}
\label{eq:sc}
\frac{\zeta}{\eta} = \kappa_\GZ \left(\frac{1}{3}-c_s^2\right),
\end{equation}
where $\kappa_\GZ = 5/2$. One should note here that the scaling (\ref{eq:sc}) is characteristic for strongly-coupled theories based on gauge-gravity duality \cite{Benincasa:2005iv,Buchel:2005cv}, and disagrees with weakly-coupled approaches \cite{Arnold:2006fz}. Equation ~(\ref{eq:sc})  is qualitatively different from the scaling found for the BE plasma where
\begin{equation}
\label{eq:sc}
\frac{\zeta}{\eta} = \kappa_\BE \left(\frac{1}{3}-c_s^2\right)^{3/2}
\end{equation}
and $\kappa_\BE = 3 \sqrt{15}/2$. On the other hand, at low temperatures we get $\zeta/\eta =5/3$ and $\zeta/\eta =2/3$ for GZ and BE plasma, respectively.
%
%
%%%%%%%%%%%%%%%%%%%%%%%%%%%%%%%%%%%%%%%%%%%%%%%%%%%%%%%%%%%%%%%%%%%%%%%%%%
\section{Conclusions}
\label{s-conc}
%%%%%%%%%%%%%%%%%%%%%%%%%%%%%%%%%%%%%%%%%%%%%%%%%%%%%%%%%%%%%%%%%%%%%%%%%%
%
In this proceedings contribution we have reviewed our main results on the non-equilibrium properties of the Gribov-Zwanziger plasma of confining gluons \cite{Gribov:1977wm,Zwanziger:1989mf} found in Refs.~\cite{Florkowski:2015rua,Florkowski:2015dmm}. In particular, we have emphasized certain features of the bulk and shear viscosities, such as the temperature dependence of their ratio, which are characteristics of the strongly-coupled theories \cite{Benincasa:2005iv,Buchel:2005cv}. These results were confronted with those obtained for the non-interacting massive Bose-Einstein plasma. 
%%%%%%%%%%%%%%%%%%%%%%%%%%%%%%%%%%%%%%%%%%%%%%%%%%%%%%%%%%%%%%%%%%%%%%%%%%
\section*{Acknowledgements}
\label{s-ack}
%%%%%%%%%%%%%%%%%%%%%%%%%%%%%%%%%%%%%%%%%%%%%%%%%%%%%%%%%%%%%%%%%%%%%%%%%%
\par Research  supported  in  part  by  Polish  National  Science  Center  grants  No. DEC-2012/05/B/ST2/02528, No.  DEC-2012/06/A/ST2/00390 (W.F.) and No.
DEC-2012/07/D/ST2/02125 (R.R.). N.S. was supported by the Helmholtz International Center for FAIR within the framework of the LOEWE program launched by the State of Hesse. K.T. was supported by a Juan de la Cierva fellowship and by the Spanish MINECO under projects  FPA2013-46570 and 2014SGR104, partially by MDM-2014-0369 of ICCUB (Unidad de Excelencia 'Mar\'ia de Maeztu'), by the Consolider CPAN project and by FEDER.
%%%%%%%%%%%%%%%%%%%%%%%%%%%%%%%%%%%%%%%%%%%%%%%%%%%%%%%%%%%%%%%%%%%%%%%%%%
\bibliographystyle{elsarticle-num}
\bibliography{refs}

\begin{thebibliography}{10}
\expandafter\ifx\csname url\endcsname\relax
  \def\url#1{\texttt{#1}}\fi
\expandafter\ifx\csname urlprefix\endcsname\relax\def\urlprefix{URL }\fi
\expandafter\ifx\csname href\endcsname\relax
  \def\href#1#2{#2} \def\path#1{#1}\fi

\bibitem{Borsanyi:2013bia}
S.~Borsanyi, Z.~Fodor, C.~Hoelbling, S.~D. Katz, S.~Krieg, K.~K. Szabo, {Full
  result for the QCD equation of state with 2+1 flavors}, Phys. Lett. B730
  (2014) 99--104.
\newblock \href {http://arxiv.org/abs/1309.5258} {\path{arXiv:1309.5258}},
  \href {http://dx.doi.org/10.1016/j.physletb.2014.01.007}
  {\path{doi:10.1016/j.physletb.2014.01.007}}.

\bibitem{Asakawa:2013laa}
M.~Asakawa, T.~Hatsuda, E.~Itou, M.~Kitazawa, H.~Suzuki, {Thermodynamics of
  SU(3) gauge theory from gradient flow on the lattice}, Phys. Rev. D90~(1)
  (2014) 011501, [Erratum: Phys. Rev.D92,no.5,059902(2015)].
\newblock \href {http://arxiv.org/abs/1312.7492} {\path{arXiv:1312.7492}},
  \href {http://dx.doi.org/10.1103/PhysRevD.90.011501,
  10.1103/PhysRevD.92.059902} {\path{doi:10.1103/PhysRevD.90.011501,
  10.1103/PhysRevD.92.059902}}.

\bibitem{Bazavov:2014pvz}
A.~Bazavov, et~al., {Equation of state in ( 2+1 )-flavor QCD}, Phys. Rev.
  D90~(9) (2014) 094503.
\newblock \href {http://arxiv.org/abs/1407.6387} {\path{arXiv:1407.6387}},
  \href {http://dx.doi.org/10.1103/PhysRevD.90.094503}
  {\path{doi:10.1103/PhysRevD.90.094503}}.

\bibitem{Andersen:2011sf}
J.~O. Andersen, L.~E. Leganger, M.~Strickland, N.~Su, {Three-loop HTL QCD
  thermodynamics}, JHEP 08 (2011) 053.
\newblock \href {http://arxiv.org/abs/1103.2528} {\path{arXiv:1103.2528}},
  \href {http://dx.doi.org/10.1007/JHEP08(2011)053}
  {\path{doi:10.1007/JHEP08(2011)053}}.

\bibitem{Mogliacci:2013mca}
S.~Mogliacci, J.~O. Andersen, M.~Strickland, N.~Su, A.~Vuorinen, {Equation of
  State of hot and dense QCD: Resummed perturbation theory confronts lattice
  data}, JHEP 12 (2013) 055.
\newblock \href {http://arxiv.org/abs/1307.8098} {\path{arXiv:1307.8098}},
  \href {http://dx.doi.org/10.1007/JHEP12(2013)055}
  {\path{doi:10.1007/JHEP12(2013)055}}.

\bibitem{Haque:2014rua}
N.~Haque, A.~Bandyopadhyay, J.~O. Andersen, M.~G. Mustafa, M.~Strickland,
  N.~Su, {Three-loop HTLpt thermodynamics at finite temperature and chemical
  potential}, JHEP 05 (2014) 027.
\newblock \href {http://arxiv.org/abs/1402.6907} {\path{arXiv:1402.6907}},
  \href {http://dx.doi.org/10.1007/JHEP05(2014)027}
  {\path{doi:10.1007/JHEP05(2014)027}}.

\bibitem{Su:2015esa}
N.~Su, {Recent progress in hard-thermal-loop QCD thermodynamics and collective
  excitations}, Int. J. Mod. Phys. A30~(09) (2015) 1530025.
\newblock \href {http://arxiv.org/abs/1502.04589} {\path{arXiv:1502.04589}},
  \href {http://dx.doi.org/10.1142/S0217751X15300252}
  {\path{doi:10.1142/S0217751X15300252}}.

\bibitem{Meyer:2007dy}
H.~B. Meyer, {A Calculation of the bulk viscosity in SU(3) gluodynamics}, Phys.
  Rev. Lett. 100 (2008) 162001.
\newblock \href {http://arxiv.org/abs/0710.3717} {\path{arXiv:0710.3717}},
  \href {http://dx.doi.org/10.1103/PhysRevLett.100.162001}
  {\path{doi:10.1103/PhysRevLett.100.162001}}.

\bibitem{Meyer:2007ic}
H.~B. Meyer, {A Calculation of the shear viscosity in SU(3) gluodynamics},
  Phys. Rev. D76 (2007) 101701.
\newblock \href {http://arxiv.org/abs/0704.1801} {\path{arXiv:0704.1801}},
  \href {http://dx.doi.org/10.1103/PhysRevD.76.101701}
  {\path{doi:10.1103/PhysRevD.76.101701}}.

\bibitem{Florkowski:2015rua}
W.~Florkowski, R.~Ryblewski, N.~Su, K.~Tywoniuk, {Bulk viscosity in a plasma of
  confining gluons}\href {http://arxiv.org/abs/1504.03176}
  {\path{arXiv:1504.03176}}.

\bibitem{Florkowski:2015dmm}
W.~Florkowski, R.~Ryblewski, N.~Su, K.~Tywoniuk, {Transport coefficients of the
  Gribov-Zwanziger plasma}\href {http://arxiv.org/abs/1509.01242}
  {\path{arXiv:1509.01242}}.

\bibitem{Gribov:1977wm}
V.~N. Gribov, {Quantization of Nonabelian Gauge Theories}, Nucl. Phys. B139
  (1978) 1.
\newblock \href {http://dx.doi.org/10.1016/0550-3213(78)90175-X}
  {\path{doi:10.1016/0550-3213(78)90175-X}}.

\bibitem{Zwanziger:1989mf}
D.~Zwanziger, {Local and Renormalizable Action From the Gribov Horizon}, Nucl.
  Phys. B323 (1989) 513--544.
\newblock \href {http://dx.doi.org/10.1016/0550-3213(89)90122-3}
  {\path{doi:10.1016/0550-3213(89)90122-3}}.

\bibitem{Zwanziger:2004np}
D.~Zwanziger, {Equation of state of gluon plasma from fundamental modular
  region}, Phys. Rev. Lett. 94 (2005) 182301.
\newblock \href {http://arxiv.org/abs/hep-ph/0407103}
  {\path{arXiv:hep-ph/0407103}}, \href
  {http://dx.doi.org/10.1103/PhysRevLett.94.182301}
  {\path{doi:10.1103/PhysRevLett.94.182301}}.

\bibitem{Zwanziger:2006sc}
D.~Zwanziger, {Equation of State of Gluon Plasma from Local Action}, Phys. Rev.
  D76 (2007) 125014.
\newblock \href {http://arxiv.org/abs/hep-ph/0610021}
  {\path{arXiv:hep-ph/0610021}}, \href
  {http://dx.doi.org/10.1103/PhysRevD.76.125014}
  {\path{doi:10.1103/PhysRevD.76.125014}}.

\bibitem{Kojo:2009ha}
T.~Kojo, Y.~Hidaka, L.~McLerran, R.~D. Pisarski, {Quarkyonic Chiral Spirals},
  Nucl. Phys. A843 (2010) 37--58.
\newblock \href {http://arxiv.org/abs/0912.3800} {\path{arXiv:0912.3800}},
  \href {http://dx.doi.org/10.1016/j.nuclphysa.2010.05.053}
  {\path{doi:10.1016/j.nuclphysa.2010.05.053}}.

\bibitem{Kojo:2012js}
T.~Kojo, N.~Su, {The quark mass gap in a magnetic field}, Phys. Lett. B720
  (2013) 192--197.
\newblock \href {http://arxiv.org/abs/1211.7318} {\path{arXiv:1211.7318}},
  \href {http://dx.doi.org/10.1016/j.physletb.2013.02.024}
  {\path{doi:10.1016/j.physletb.2013.02.024}}.

\bibitem{Fukushima:2013xsa}
K.~Fukushima, N.~Su, {Stabilizing perturbative Yang-Mills thermodynamics with
  Gribov quantization}, Phys.Rev. D88 (2013) 076008.
\newblock \href {http://arxiv.org/abs/1304.8004} {\path{arXiv:1304.8004}},
  \href {http://dx.doi.org/10.1103/PhysRevD.88.076008}
  {\path{doi:10.1103/PhysRevD.88.076008}}.

\bibitem{Su:2014rma}
N.~Su, K.~Tywoniuk, {Massless Mode and Positivity Violation in Hot QCD}, Phys.
  Rev. Lett. 114~(16) (2015) 161601.
\newblock \href {http://arxiv.org/abs/1409.3203} {\path{arXiv:1409.3203}},
  \href {http://dx.doi.org/10.1103/PhysRevLett.114.161601}
  {\path{doi:10.1103/PhysRevLett.114.161601}}.

\bibitem{Kharzeev:2015xsa}
D.~E. Kharzeev, E.~M. Levin, {Color Confinement and Screening in the $\theta$
  Vacuum of QCD}, Phys. Rev. Lett. 114~(24) (2015) 242001.
\newblock \href {http://arxiv.org/abs/1501.04622} {\path{arXiv:1501.04622}},
  \href {http://dx.doi.org/10.1103/PhysRevLett.114.242001}
  {\path{doi:10.1103/PhysRevLett.114.242001}}.

\bibitem{Canfora:2015yia}
F.~E. Canfora, D.~Dudal, I.~F. Justo, P.~Pais, L.~Rosa, D.~Vercauteren, {Effect
  of the Gribov horizon on the Polyakov loop and vice versa}, Eur. Phys. J.
  C75~(7) (2015) 326.
\newblock \href {http://arxiv.org/abs/1505.02287} {\path{arXiv:1505.02287}},
  \href {http://dx.doi.org/10.1140/epjc/s10052-015-3546-y}
  {\path{doi:10.1140/epjc/s10052-015-3546-y}}.

\bibitem{Bandyopadhyay:2015wua}
A.~Bandyopadhyay, N.~Haque, M.~G. Mustafa, M.~Strickland, {Dilepton rate and
  quark number susceptibility with the Gribov action}\href
  {http://arxiv.org/abs/1508.06249} {\path{arXiv:1508.06249}}.

\bibitem{Bjorken:1982qr}
J.~D. Bjorken, {Highly Relativistic Nucleus-Nucleus Collisions: The Central
  Rapidity Region}, Phys. Rev. D27 (1983) 140--151.
\newblock \href {http://dx.doi.org/10.1103/PhysRevD.27.140}
  {\path{doi:10.1103/PhysRevD.27.140}}.

\bibitem{Florkowski:2013lya}
W.~Florkowski, R.~Ryblewski, M.~Strickland, {Testing viscous and anisotropic
  hydrodynamics in an exactly solvable case}, Phys.Rev. C88 (2013) 024903.
\newblock \href {http://arxiv.org/abs/1305.7234} {\path{arXiv:1305.7234}},
  \href {http://dx.doi.org/10.1103/PhysRevC.88.024903}
  {\path{doi:10.1103/PhysRevC.88.024903}}.

\bibitem{Sanches:2015vra}
S.~M. Sanches, D.~A. Fogaça, F.~S. Navarra, H.~Marrochio, {Cavitation in a
  quark gluon plasma with finite chemical potential and several transport
  coefficients}, Phys. Rev. C92~(2) (2015) 025204.
\newblock \href {http://arxiv.org/abs/1505.06335} {\path{arXiv:1505.06335}},
  \href {http://dx.doi.org/10.1103/PhysRevC.92.025204}
  {\path{doi:10.1103/PhysRevC.92.025204}}.

\bibitem{Benincasa:2005iv}
P.~Benincasa, A.~Buchel, A.~O. Starinets, {Sound waves in strongly coupled
  non-conformal gauge theory plasma}, Nucl. Phys. B733 (2006) 160--187.
\newblock \href {http://arxiv.org/abs/hep-th/0507026}
  {\path{arXiv:hep-th/0507026}}, \href
  {http://dx.doi.org/10.1016/j.nuclphysb.2005.11.005}
  {\path{doi:10.1016/j.nuclphysb.2005.11.005}}.

\bibitem{Buchel:2005cv}
A.~Buchel, {Transport properties of cascading gauge theories}, Phys. Rev. D72
  (2005) 106002.
\newblock \href {http://arxiv.org/abs/hep-th/0509083}
  {\path{arXiv:hep-th/0509083}}, \href
  {http://dx.doi.org/10.1103/PhysRevD.72.106002}
  {\path{doi:10.1103/PhysRevD.72.106002}}.

\bibitem{Arnold:2006fz}
P.~B. Arnold, C.~Dogan, G.~D. Moore, {The Bulk Viscosity of High-Temperature
  QCD}, Phys. Rev. D74 (2006) 085021.
\newblock \href {http://arxiv.org/abs/hep-ph/0608012}
  {\path{arXiv:hep-ph/0608012}}, \href
  {http://dx.doi.org/10.1103/PhysRevD.74.085021}
  {\path{doi:10.1103/PhysRevD.74.085021}}.

\end{thebibliography}


\providecommand{\href}[2]{#2}\begingroup\raggedright\begin{thebibliography}{10}

\bibitem{ALICE:2011ab}
{\bf ALICE} Collaboration, K.~Aamodt {\em et.~al.}, {\it {Higher harmonic
  anisotropic flow measurements of charged particles in Pb-Pb collisions at
  $\sqrt{s\_{NN}}$=2.76 TeV}},  {\em Phys. Rev. Lett.} {\bf 107} (2011) 032301,
  [\href{http://xxx.lanl.gov/abs/1105.3865}{{\tt arXiv:1105.3865}}].

\bibitem{Aamodt:2011by}
{\bf ALICE} Collaboration, K.~Aamodt {\em et.~al.}, {\it {Harmonic
  decomposition of two-particle angular correlations in Pb-Pb collisions at
  $\sqrt{s\_{NN}}=2.76$ TeV}},  {\em Phys. Lett.} {\bf B708} (2012) 249--264,
  [\href{http://xxx.lanl.gov/abs/1109.2501}{{\tt arXiv:1109.2501}}].

\bibitem{Chatrchyan:2012wg}
{\bf CMS} Collaboration, S.~Chatrchyan {\em et.~al.}, {\it {Centrality
  dependence of dihadron correlations and azimuthal anisotropy harmonics in
  PbPb collisions at $\sqrt{s\_{NN}}=2.76$ TeV}},  {\em Eur. Phys. J.} {\bf
  C72} (2012) 2012, [\href{http://xxx.lanl.gov/abs/1201.3158}{{\tt
  arXiv:1201.3158}}].

\bibitem{Aad:2013xma}
{\bf ATLAS} Collaboration, G.~Aad {\em et.~al.}, {\it {Measurement of the
  distributions of event-by-event flow harmonics in lead-lead collisions at =
  2.76 TeV with the ATLAS detector at the LHC}},  {\em JHEP} {\bf 1311} (2013)
  183, [\href{http://xxx.lanl.gov/abs/1305.2942}{{\tt arXiv:1305.2942}}].

\bibitem{Adamczyk:2013waa}
{\bf STAR} Collaboration, L.~Adamczyk {\em et.~al.}, {\it {Third Harmonic Flow
  of Charged Particles in Au+Au Collisions at sqrtsNN = 200 GeV}},  {\em
  Phys.Rev.} {\bf C88} (2013), no.~1 014904,
  [\href{http://xxx.lanl.gov/abs/1301.2187}{{\tt arXiv:1301.2187}}].

\bibitem{Adare:2011tg}
{\bf PHENIX} Collaboration, A.~Adare {\em et.~al.}, {\it {Measurements of
  Higher-Order Flow Harmonics in Au+Au Collisions at $\sqrt{s\_{NN}} = 200$
  GeV}},  {\em Phys. Rev. Lett.} {\bf 107} (2011) 252301,
  [\href{http://xxx.lanl.gov/abs/1105.3928}{{\tt arXiv:1105.3928}}].

\bibitem{Gyulassy:2004zy}
M.~Gyulassy and L.~McLerran, {\it {New forms of QCD matter discovered at
  RHIC}},  {\em Nucl. Phys.} {\bf A750} (2005) 30--63,
  [\href{http://xxx.lanl.gov/abs/nucl-th/0405013}{{\tt nucl-th/0405013}}].

\bibitem{Shuryak:2004cy}
E.~V. Shuryak, {\it {What RHIC experiments and theory tell us about properties
  of quark-gluon plasma?}},  {\em Nucl. Phys.} {\bf A750} (2005) 64--83,
  [\href{http://xxx.lanl.gov/abs/hep-ph/0405066}{{\tt hep-ph/0405066}}].

\bibitem{Romatschke:2009kr}
P.~Romatschke, {\it {Relativistic Viscous Fluid Dynamics and Non-Equilibrium
  Entropy}},  {\em Class. Quant. Grav.} {\bf 27} (2010) 025006,
  [\href{http://xxx.lanl.gov/abs/0906.4787}{{\tt arXiv:0906.4787}}].

\bibitem{Jeon:2015dfa}
S.~Jeon and U.~Heinz, {\it {Introduction to Hydrodynamics}},
  \href{http://xxx.lanl.gov/abs/1503.0393}{{\tt arXiv:1503.0393}}.

\bibitem{Gale:2013da}
C.~Gale, S.~Jeon, and B.~Schenke, {\it {Hydrodynamic Modeling of Heavy-Ion
  Collisions}},  {\em Int.J.Mod.Phys.} {\bf A28} (2013) 1340011,
  [\href{http://xxx.lanl.gov/abs/1301.5893}{{\tt arXiv:1301.5893}}].

\bibitem{Jeon:1995zm}
S.~Jeon and L.~G. Yaffe, {\it {From quantum field theory to hydrodynamics:
  Transport coefficients and effective kinetic theory}},  {\em Phys. Rev.} {\bf
  D53} (1996) 5799--5809, [\href{http://xxx.lanl.gov/abs/hep-ph/9512263}{{\tt
  hep-ph/9512263}}].

\bibitem{Andersen:2004fp}
J.~O. Andersen and M.~Strickland, {\it {Resummation in hot field theories}},
  {\em Annals Phys.} {\bf 317} (2005) 281--353,
  [\href{http://xxx.lanl.gov/abs/hep-ph/0404164}{{\tt hep-ph/0404164}}].

\bibitem{Su:2012iy}
N.~Su, {\it {A brief overview of hard-thermal-loop perturbation theory}},  {\em
  Commun. Theor. Phys.} {\bf 57} (2012) 409,
  [\href{http://xxx.lanl.gov/abs/1204.0260}{{\tt arXiv:1204.0260}}].

\bibitem{Su:2015esa}
N.~Su, {\it {Recent progress in hard-thermal-loop QCD thermodynamics and
  collective excitations}},  {\em Int. J. Mod. Phys.} {\bf A30} (2015), no.~09
  1530025, [\href{http://xxx.lanl.gov/abs/1502.0458}{{\tt arXiv:1502.0458}}].

\bibitem{Andersen:2010ct}
J.~O. Andersen, M.~Strickland, and N.~Su, {\it {Three-loop HTL gluon
  thermodynamics at intermediate coupling}},  {\em JHEP} {\bf 08} (2010) 113,
  [\href{http://xxx.lanl.gov/abs/1005.1603}{{\tt arXiv:1005.1603}}].

\bibitem{Andersen:2011sf}
J.~O. Andersen, L.~E. Leganger, M.~Strickland, and N.~Su, {\it {Three-loop HTL
  QCD thermodynamics}},  {\em JHEP} {\bf 08} (2011) 053,
  [\href{http://xxx.lanl.gov/abs/1103.2528}{{\tt arXiv:1103.2528}}].

\bibitem{Mogliacci:2013mca}
S.~Mogliacci, J.~O. Andersen, M.~Strickland, N.~Su, and A.~Vuorinen, {\it
  {Equation of State of hot and dense QCD: Resummed perturbation theory
  confronts lattice data}},  {\em JHEP} {\bf 12} (2013) 055,
  [\href{http://xxx.lanl.gov/abs/1307.8098}{{\tt arXiv:1307.8098}}].

\bibitem{Haque:2014rua}
N.~Haque, A.~Bandyopadhyay, J.~O. Andersen, M.~G. Mustafa, M.~Strickland, and
  N.~Su, {\it {Three-loop HTLpt thermodynamics at finite temperature and
  chemical potential}},  {\em JHEP} {\bf 05} (2014) 027,
  [\href{http://xxx.lanl.gov/abs/1402.6907}{{\tt arXiv:1402.6907}}].

\bibitem{Bazavov:2014pvz}
{\bf HotQCD} Collaboration, A.~Bazavov {\em et.~al.}, {\it {Equation of state
  in ( 2+1 )-flavor QCD}},  {\em Phys. Rev.} {\bf D90} (2014), no.~9 094503,
  [\href{http://xxx.lanl.gov/abs/1407.6387}{{\tt arXiv:1407.6387}}].

\bibitem{Borsanyi:2013bia}
S.~Borsanyi, Z.~Fodor, C.~Hoelbling, S.~D. Katz, S.~Krieg, and K.~K. Szabo,
  {\it {Full result for the QCD equation of state with 2+1 flavors}},  {\em
  Phys. Lett.} {\bf B730} (2014) 99--104,
  [\href{http://xxx.lanl.gov/abs/1309.5258}{{\tt arXiv:1309.5258}}].

\bibitem{Meyer:2011gj}
H.~B. Meyer, {\it {Transport Properties of the Quark-Gluon Plasma: A Lattice
  QCD Perspective}},  {\em Eur. Phys. J.} {\bf A47} (2011) 86,
  [\href{http://xxx.lanl.gov/abs/1104.3708}{{\tt arXiv:1104.3708}}].

\bibitem{Blaizot:2001nr}
J.-P. Blaizot and E.~Iancu, {\it {The Quark gluon plasma: Collective dynamics
  and hard thermal loops}},  {\em Phys. Rept.} {\bf 359} (2002) 355--528,
  [\href{http://xxx.lanl.gov/abs/hep-ph/0101103}{{\tt hep-ph/0101103}}].

\bibitem{Arnold:2002zm}
P.~B. Arnold, G.~D. Moore, and L.~G. Yaffe, {\it {Effective kinetic theory for
  high temperature gauge theories}},  {\em JHEP} {\bf 01} (2003) 030,
  [\href{http://xxx.lanl.gov/abs/hep-ph/0209353}{{\tt hep-ph/0209353}}].

\bibitem{Hong:2010at}
J.~Hong and D.~Teaney, {\it {Spectral densities for hot QCD plasmas in a
  leading log approximation}},  {\em Phys. Rev.} {\bf C82} (2010) 044908,
  [\href{http://xxx.lanl.gov/abs/1003.0699}{{\tt arXiv:1003.0699}}].

\bibitem{Moore:2008ws}
G.~D. Moore and O.~Saremi, {\it {Bulk viscosity and spectral functions in
  QCD}},  {\em JHEP} {\bf 09} (2008) 015,
  [\href{http://xxx.lanl.gov/abs/0805.4201}{{\tt arXiv:0805.4201}}].

\bibitem{Florkowski:2010cf}
W.~Florkowski and R.~Ryblewski, {\it {Highly-anisotropic and
  strongly-dissipative hydrodynamics for early stages of relativistic heavy-ion
  collisions}},  {\em Phys. Rev.} {\bf C83} (2011) 034907,
  [\href{http://xxx.lanl.gov/abs/1007.0130}{{\tt arXiv:1007.0130}}].

\bibitem{Martinez:2010sc}
M.~Martinez and M.~Strickland, {\it {Dissipative Dynamics of Highly Anisotropic
  Systems}},  {\em Nucl. Phys.} {\bf A848} (2010) 183--197,
  [\href{http://xxx.lanl.gov/abs/1007.0889}{{\tt arXiv:1007.0889}}].

\bibitem{Strickland:2014pga}
M.~Strickland, {\it {Anisotropic Hydrodynamics: Three lectures}},  {\em Acta
  Phys. Polon.} {\bf B45} (2014), no.~12 2355--2394,
  [\href{http://xxx.lanl.gov/abs/1410.5786}{{\tt arXiv:1410.5786}}].

\bibitem{CasalderreySolana:2011us}
J.~Casalderrey-Solana, H.~Liu, D.~Mateos, K.~Rajagopal, and U.~A. Wiedemann,
  {\it {Gauge/String Duality, Hot QCD and Heavy Ion Collisions}},
  \href{http://xxx.lanl.gov/abs/1101.0618}{{\tt arXiv:1101.0618}}.

\bibitem{Kovtun:2006pf}
P.~Kovtun and A.~Starinets, {\it {Thermal spectral functions of strongly
  coupled N=4 supersymmetric Yang-Mills theory}},  {\em Phys. Rev. Lett.} {\bf
  96} (2006) 131601, [\href{http://xxx.lanl.gov/abs/hep-th/0602059}{{\tt
  hep-th/0602059}}].

\bibitem{Teaney:2006nc}
D.~Teaney, {\it {Finite temperature spectral densities of momentum and R-charge
  correlators in N=4 Yang Mills theory}},  {\em Phys. Rev.} {\bf D74} (2006)
  045025, [\href{http://xxx.lanl.gov/abs/hep-ph/0602044}{{\tt
  hep-ph/0602044}}].

\bibitem{Heller:2014wfa}
M.~P. Heller, R.~A. Janik, M.~Spalinski, and P.~Witaszczyk, {\it {Coupling
  hydrodynamics to nonequilibrium degrees of freedom in strongly interacting
  quark-gluon plasma}},  {\em Phys. Rev. Lett.} {\bf 113} (2014) 261601,
  [\href{http://xxx.lanl.gov/abs/1409.5087}{{\tt arXiv:1409.5087}}].

\bibitem{Arnold:2000dr}
P.~B. Arnold, G.~D. Moore, and L.~G. Yaffe, {\it {Transport coefficients in
  high temperature gauge theories. 1. Leading log results}},  {\em JHEP} {\bf
  11} (2000) 001, [\href{http://xxx.lanl.gov/abs/hep-ph/0010177}{{\tt
  hep-ph/0010177}}].

\bibitem{Arnold:2003zc}
P.~B. Arnold, G.~D. Moore, and L.~G. Yaffe, {\it {Transport coefficients in
  high temperature gauge theories. 2. Beyond leading log}},  {\em JHEP} {\bf
  05} (2003) 051, [\href{http://xxx.lanl.gov/abs/hep-ph/0302165}{{\tt
  hep-ph/0302165}}].

\bibitem{Arnold:2006fz}
P.~B. Arnold, C.~Dogan, and G.~D. Moore, {\it {The Bulk Viscosity of
  High-Temperature QCD}},  {\em Phys. Rev.} {\bf D74} (2006) 085021,
  [\href{http://xxx.lanl.gov/abs/hep-ph/0608012}{{\tt hep-ph/0608012}}].

\bibitem{Meyer:2007dy}
H.~B. Meyer, {\it {A Calculation of the bulk viscosity in SU(3) gluodynamics}},
   {\em Phys. Rev. Lett.} {\bf 100} (2008) 162001,
  [\href{http://xxx.lanl.gov/abs/0710.3717}{{\tt arXiv:0710.3717}}].

\bibitem{Meyer:2007ic}
H.~B. Meyer, {\it {A Calculation of the shear viscosity in SU(3)
  gluodynamics}},  {\em Phys. Rev.} {\bf D76} (2007) 101701,
  [\href{http://xxx.lanl.gov/abs/0704.1801}{{\tt arXiv:0704.1801}}].

\bibitem{Kharzeev:2007wb}
D.~Kharzeev and K.~Tuchin, {\it {Bulk viscosity of QCD matter near the critical
  temperature}},  {\em JHEP} {\bf 09} (2008) 093,
  [\href{http://xxx.lanl.gov/abs/0705.4280}{{\tt arXiv:0705.4280}}].

\bibitem{Karsch:2007jc}
F.~Karsch, D.~Kharzeev, and K.~Tuchin, {\it {Universal properties of bulk
  viscosity near the QCD phase transition}},  {\em Phys. Lett.} {\bf B663}
  (2008) 217--221, [\href{http://xxx.lanl.gov/abs/0711.0914}{{\tt
  arXiv:0711.0914}}].

\bibitem{Haas:2013hpa}
M.~Haas, L.~Fister, and J.~M. Pawlowski, {\it {Gluon spectral functions and
  transport coefficients in Yang--Mills theory}},  {\em Phys. Rev.} {\bf D90}
  (2014), no.~9 091501, [\href{http://xxx.lanl.gov/abs/1308.4960}{{\tt
  arXiv:1308.4960}}].

\bibitem{Christiansen:2014ypa}
N.~Christiansen, M.~Haas, J.~M. Pawlowski, and N.~Strodthoff, {\it {Transport
  Coefficients in Yang--Mills Theory and QCD}},
  \href{http://xxx.lanl.gov/abs/1411.7986}{{\tt arXiv:1411.7986}}.

\bibitem{Policastro:2001yc}
G.~Policastro, D.~T. Son, and A.~O. Starinets, {\it {The Shear viscosity of
  strongly coupled N=4 supersymmetric Yang-Mills plasma}},  {\em Phys. Rev.
  Lett.} {\bf 87} (2001) 081601,
  [\href{http://xxx.lanl.gov/abs/hep-th/0104066}{{\tt hep-th/0104066}}].

\bibitem{Benincasa:2005iv}
P.~Benincasa, A.~Buchel, and A.~O. Starinets, {\it {Sound waves in strongly
  coupled non-conformal gauge theory plasma}},  {\em Nucl. Phys.} {\bf B733}
  (2006) 160--187, [\href{http://xxx.lanl.gov/abs/hep-th/0507026}{{\tt
  hep-th/0507026}}].

\bibitem{Buchel:2005cv}
A.~Buchel, {\it {Transport properties of cascading gauge theories}},  {\em
  Phys. Rev.} {\bf D72} (2005) 106002,
  [\href{http://xxx.lanl.gov/abs/hep-th/0509083}{{\tt hep-th/0509083}}].

\bibitem{Finazzo:2014cna}
S.~I. Finazzo, R.~Rougemont, H.~Marrochio, and J.~Noronha, {\it {Hydrodynamic
  transport coefficients for the non-conformal quark-gluon plasma from
  holography}},  {\em JHEP} {\bf 02} (2015) 051,
  [\href{http://xxx.lanl.gov/abs/1412.2968}{{\tt arXiv:1412.2968}}].

\bibitem{Li:2014dsa}
D.~Li, S.~He, and M.~Huang, {\it {Temperature dependent transport coefficients
  in a dynamical holographic QCD model}},  {\em JHEP} {\bf 06} (2015) 046,
  [\href{http://xxx.lanl.gov/abs/1411.5332}{{\tt arXiv:1411.5332}}].

\bibitem{Linde:1980ts}
A.~D. Linde, {\it {Infrared Problem in Thermodynamics of the Yang-Mills Gas}},
  {\em Phys. Lett.} {\bf B96} (1980) 289.

\bibitem{Gross:1980br}
D.~J. Gross, R.~D. Pisarski, and L.~G. Yaffe, {\it {QCD and Instantons at
  Finite Temperature}},  {\em Rev. Mod. Phys.} {\bf 53} (1981) 43.

\bibitem{Gribov:1977wm}
V.~N. Gribov, {\it {Quantization of Nonabelian Gauge Theories}},  {\em Nucl.
  Phys.} {\bf B139} (1978) 1.

\bibitem{Zwanziger:1989mf}
D.~Zwanziger, {\it {Local and Renormalizable Action From the Gribov Horizon}},
  {\em Nucl. Phys.} {\bf B323} (1989) 513--544.

\bibitem{Dokshitzer:2004ie}
Y.~L. Dokshitzer and D.~E. Kharzeev, {\it {The Gribov conception of quantum
  chromodynamics}},  {\em Ann. Rev. Nucl. Part. Sci.} {\bf 54} (2004) 487--524,
  [\href{http://xxx.lanl.gov/abs/hep-ph/0404216}{{\tt hep-ph/0404216}}].

\bibitem{Vandersickel:2012tz}
N.~Vandersickel and D.~Zwanziger, {\it {The Gribov problem and QCD dynamics}},
  {\em Phys. Rept.} {\bf 520} (2012) 175--251,
  [\href{http://xxx.lanl.gov/abs/1202.1491}{{\tt arXiv:1202.1491}}].

\bibitem{Feynman:1981ss}
R.~P. Feynman, {\it {The Qualitative Behavior of Yang-Mills Theory in
  (2+1)-Dimensions}},  {\em Nucl. Phys.} {\bf B188} (1981) 479.

\bibitem{Burgio:2008jr}
G.~Burgio, M.~Quandt, and H.~Reinhardt, {\it {Coulomb gauge gluon propagator
  and the Gribov formula}},  {\em Phys. Rev. Lett.} {\bf 102} (2009) 032002,
  [\href{http://xxx.lanl.gov/abs/0807.3291}{{\tt arXiv:0807.3291}}].

\bibitem{Alkofer:2000wg}
R.~Alkofer and L.~von Smekal, {\it {The Infrared behavior of QCD Green's
  functions: Confinement dynamical symmetry breaking, and hadrons as
  relativistic bound states}},  {\em Phys. Rept.} {\bf 353} (2001) 281,
  [\href{http://xxx.lanl.gov/abs/hep-ph/0007355}{{\tt hep-ph/0007355}}].

\bibitem{Maas:2011se}
A.~Maas, {\it {Describing gauge bosons at zero and finite temperature}},  {\em
  Phys. Rept.} {\bf 524} (2013) 203--300,
  [\href{http://xxx.lanl.gov/abs/1106.3942}{{\tt arXiv:1106.3942}}].

\bibitem{Zwanziger:2004np}
D.~Zwanziger, {\it {Equation of state of gluon plasma from fundamental modular
  region}},  {\em Phys. Rev. Lett.} {\bf 94} (2005) 182301,
  [\href{http://xxx.lanl.gov/abs/hep-ph/0407103}{{\tt hep-ph/0407103}}].

\bibitem{Fukushima:2013xsa}
K.~Fukushima and N.~Su, {\it {Stabilizing perturbative Yang-Mills
  thermodynamics with Gribov quantization}},  {\em Phys.Rev.} {\bf D88} (2013)
  076008, [\href{http://xxx.lanl.gov/abs/1304.8004}{{\tt arXiv:1304.8004}}].

\bibitem{Zwanziger:2006sc}
D.~Zwanziger, {\it {Equation of State of Gluon Plasma from Local Action}},
  {\em Phys. Rev.} {\bf D76} (2007) 125014,
  [\href{http://xxx.lanl.gov/abs/hep-ph/0610021}{{\tt hep-ph/0610021}}].

\bibitem{Su:2014rma}
N.~Su and K.~Tywoniuk, {\it {Massless Mode and Positivity Violation in Hot
  QCD}},  {\em Phys. Rev. Lett.} {\bf 114} (2015), no.~16 161601,
  [\href{http://xxx.lanl.gov/abs/1409.3203}{{\tt arXiv:1409.3203}}].

\bibitem{Chernodub:2007rn}
M.~N. Chernodub and V.~I. Zakharov, {\it {Combining infrared and
  low-temperature asymptotes in Yang-Mills theories}},  {\em Phys. Rev. Lett.}
  {\bf 100} (2008) 222001, [\href{http://xxx.lanl.gov/abs/hep-ph/0703167}{{\tt
  hep-ph/0703167}}].

\bibitem{Kharzeev:2015xsa}
D.~E. Kharzeev and E.~M. Levin, {\it {Color Confinement and Screening in the
  $\theta$ Vacuum of QCD}},  {\em Phys. Rev. Lett.} {\bf 114} (2015), no.~24
  242001, [\href{http://xxx.lanl.gov/abs/1501.0462}{{\tt arXiv:1501.0462}}].

\bibitem{Canfora:2015yia}
F.~E. Canfora, D.~Dudal, I.~F. Justo, P.~Pais, L.~Rosa, and D.~Vercauteren,
  {\it {Effect of the Gribov horizon on the Polyakov loop and vice versa}},
  {\em Eur. Phys. J.} {\bf C75} (2015), no.~7 326,
  [\href{http://xxx.lanl.gov/abs/1505.0228}{{\tt arXiv:1505.0228}}].

\bibitem{Bandyopadhyay:2015wua}
A.~Bandyopadhyay, N.~Haque, M.~G. Mustafa, and M.~Strickland, {\it {Dilepton
  rate and quark number susceptibility with the Gribov action}},
  \href{http://xxx.lanl.gov/abs/1508.0624}{{\tt arXiv:1508.0624}}.

\bibitem{Florkowski:2015rua}
W.~Florkowski, R.~Ryblewski, N.~Su, and K.~Tywoniuk, {\it {Bulk viscosity in a
  plasma of confining gluons}},  \href{http://xxx.lanl.gov/abs/1504.0317}{{\tt
  arXiv:1504.0317}}.

\bibitem{Borsanyi:2012ve}
S.~Borsanyi, G.~Endrodi, Z.~Fodor, S.~D. Katz, and K.~K. Szabo, {\it {Precision
  SU(3) lattice thermodynamics for a large temperature range}},  {\em JHEP}
  {\bf 07} (2012) 056, [\href{http://xxx.lanl.gov/abs/1204.6184}{{\tt
  arXiv:1204.6184}}].

\bibitem{Bjorken:1982qr}
J.~D. Bjorken, {\it {Highly Relativistic Nucleus-Nucleus Collisions: The
  Central Rapidity Region}},  {\em Phys. Rev.} {\bf D27} (1983) 140--151.

\bibitem{Fries:2008ts}
R.~J. Fries, B.~Muller, and A.~Schafer, {\it {Stress Tensor and Bulk Viscosity
  in Relativistic Nuclear Collisions}},  {\em Phys. Rev.} {\bf C78} (2008)
  034913, [\href{http://xxx.lanl.gov/abs/0807.4333}{{\tt arXiv:0807.4333}}].

\bibitem{Bozek:2009dw}
P.~Bozek, {\it {Bulk and shear viscosities of matter created in relativistic
  heavy-ion collisions}},  {\em Phys. Rev.} {\bf C81} (2010) 034909,
  [\href{http://xxx.lanl.gov/abs/0911.2397}{{\tt arXiv:0911.2397}}].

\bibitem{Monnai:2009ad}
A.~Monnai and T.~Hirano, {\it {Effects of Bulk Viscosity at Freezeout}},  {\em
  Phys. Rev.} {\bf C80} (2009) 054906,
  [\href{http://xxx.lanl.gov/abs/0903.4436}{{\tt arXiv:0903.4436}}].

\bibitem{Denicol:2009am}
G.~S. Denicol, T.~Kodama, T.~Koide, and P.~Mota, {\it {Effect of bulk viscosity
  on Elliptic Flow near QCD phase transition}},  {\em Phys. Rev.} {\bf C80}
  (2009) 064901, [\href{http://xxx.lanl.gov/abs/0903.3595}{{\tt
  arXiv:0903.3595}}].

\bibitem{Noronha-Hostler:2013gga}
J.~Noronha-Hostler, G.~S. Denicol, J.~Noronha, R.~P.~G. Andrade, and F.~Grassi,
  {\it {Bulk Viscosity Effects in Event-by-Event Relativistic Hydrodynamics}},
  {\em Phys.Rev.} {\bf C88} (2013) 044916,
  [\href{http://xxx.lanl.gov/abs/1305.1981}{{\tt arXiv:1305.1981}}].

\bibitem{Ryu:2015vwa}
S.~Ryu, J.~F. Paquet, C.~Shen, G.~S. Denicol, B.~Schenke, S.~Jeon, and C.~Gale,
  {\it {The importance of the bulk viscosity of QCD in ultrarelativistic
  heavy-ion collisions}},  \href{http://xxx.lanl.gov/abs/1502.0167}{{\tt
  arXiv:1502.0167}}.

\bibitem{Torrieri:2008ip}
G.~Torrieri and I.~Mishustin, {\it {Instability of Boost-invariant
  hydrodynamics with a QCD inspired bulk viscosity}},  {\em Phys. Rev.} {\bf
  C78} (2008) 021901, [\href{http://xxx.lanl.gov/abs/0805.0442}{{\tt
  arXiv:0805.0442}}].

\bibitem{Rajagopal:2009yw}
K.~Rajagopal and N.~Tripuraneni, {\it {Bulk Viscosity and Cavitation in
  Boost-Invariant Hydrodynamic Expansion}},  {\em JHEP} {\bf 03} (2010) 018,
  [\href{http://xxx.lanl.gov/abs/0908.1785}{{\tt arXiv:0908.1785}}].

\bibitem{Floerchinger:2015efa}
S.~Floerchinger and M.~Martinez, {\it {Fluid dynamic propagation of initial
  baryon number perturbations on a Bjorken flow background}},
  \href{http://xxx.lanl.gov/abs/1507.0556}{{\tt arXiv:1507.0556}}.

\bibitem{Lichtenegger:2008mh}
K.~Lichtenegger and D.~Zwanziger, {\it {Nonperturbative contributions to the
  QCD pressure}},  {\em Phys. Rev.} {\bf D78} (2008) 034038,
  [\href{http://xxx.lanl.gov/abs/0805.3804}{{\tt arXiv:0805.3804}}].

\bibitem{Bialas:1987en}
A.~Bialas, W.~Czyz, A.~Dyrek, and W.~Florkowski, {\it {Oscillations of Quark -
  Gluon Plasma Generated in Strong Color Fields}},  {\em Nucl. Phys.} {\bf
  B296} (1988) 611.

\bibitem{Muronga:2003ta}
A.~Muronga, {\it {Causal theories of dissipative relativistic fluid dynamics
  for nuclear collisions}},  {\em Phys. Rev.} {\bf C69} (2004) 034903,
  [\href{http://xxx.lanl.gov/abs/nucl-th/0309055}{{\tt nucl-th/0309055}}].

\bibitem{Baier:2006um}
R.~Baier, P.~Romatschke, and U.~A. Wiedemann, {\it {Dissipative hydrodynamics
  and heavy ion collisions}},  {\em Phys. Rev.} {\bf C73} (2006) 064903,
  [\href{http://xxx.lanl.gov/abs/hep-ph/0602249}{{\tt hep-ph/0602249}}].

\bibitem{Chojnacki:2007jc}
M.~Chojnacki and W.~Florkowski, {\it {Temperature dependence of sound velocity
  and hydrodynamics of ultra-relativistic heavy-ion collisions}},  {\em Acta
  Phys. Polon.} {\bf B38} (2007) 3249--3262,
  [\href{http://xxx.lanl.gov/abs/nucl-th/0702030}{{\tt nucl-th/0702030}}].

\bibitem{Bhatnagar:1954zz}
P.~L. Bhatnagar, E.~P. Gross, and M.~Krook, {\it {A Model for Collision
  Processes in Gases. 1. Small Amplitude Processes in Charged and Neutral
  One-Component Systems}},  {\em Phys. Rev.} {\bf 94} (1954) 511--525.

\bibitem{Baym:1984np}
G.~Baym, {\it {THERMAL EQUILIBRATION IN ULTRARELATIVISTIC HEAVY ION
  COLLISIONS}},  {\em Phys. Lett.} {\bf B138} (1984) 18--22.

\bibitem{Baym:1985tna}
G.~Baym, {\it {ENTROPY PRODUCTION AND THE EVOLUTION OF ULTRARELATIVISTIC HEAVY
  ION COLLISIONS}},  {\em Nucl. Phys.} {\bf A418} (1984) 525C--537C.

\bibitem{Florkowski:2013lza}
W.~Florkowski, R.~Ryblewski, and M.~Strickland, {\it {Anisotropic Hydrodynamics
  for Rapidly Expanding Systems}},  {\em Nucl.Phys.} {\bf A916} (2013)
  249--259, [\href{http://xxx.lanl.gov/abs/1304.0665}{{\tt arXiv:1304.0665}}].

\bibitem{Florkowski:2013lya}
W.~Florkowski, R.~Ryblewski, and M.~Strickland, {\it {Testing viscous and
  anisotropic hydrodynamics in an exactly solvable case}},  {\em Phys.Rev.}
  {\bf C88} (2013) 024903, [\href{http://xxx.lanl.gov/abs/1305.7234}{{\tt
  arXiv:1305.7234}}].

\bibitem{Florkowski:2014sfa}
W.~Florkowski, E.~Maksymiuk, R.~Ryblewski, and M.~Strickland, {\it {Exact
  solution of the (0+1)-dimensional Boltzmann equation for a massive gas}},
  {\em Phys. Rev.} {\bf C89} (2014), no.~5 054908,
  [\href{http://xxx.lanl.gov/abs/1402.7348}{{\tt arXiv:1402.7348}}].

\bibitem{Weinberg:1971mx}
S.~Weinberg, {\it {Entropy generation and the survival of protogalaxies in an
  expanding universe}},  {\em Astrophys. J.} {\bf 168} (1971) 175.

\bibitem{Florkowski:2015lra}
W.~Florkowski, A.~Jaiswal, E.~Maksymiuk, R.~Ryblewski, and M.~Strickland, {\it
  {Relativistic quantum transport coefficients for second-order viscous
  hydrodynamics}},  {\em Phys. Rev.} {\bf C91} (2015) 054907,
  [\href{http://xxx.lanl.gov/abs/1503.0322}{{\tt arXiv:1503.0322}}].

\bibitem{Kovtun:2005ev}
P.~K. Kovtun and A.~O. Starinets, {\it {Quasinormal modes and holography}},
  {\em Phys. Rev.} {\bf D72} (2005) 086009,
  [\href{http://xxx.lanl.gov/abs/hep-th/0506184}{{\tt hep-th/0506184}}].

\bibitem{Sasaki:2008fg}
C.~Sasaki and K.~Redlich, {\it {Bulk viscosity in quasi particle models}},
  {\em Phys. Rev.} {\bf C79} (2009) 055207,
  [\href{http://xxx.lanl.gov/abs/0806.4745}{{\tt arXiv:0806.4745}}].

\bibitem{Huang:2010sa}
X.-G. Huang, T.~Kodama, T.~Koide, and D.~H. Rischke, {\it {Bulk Viscosity and
  Relaxation Time of Causal Dissipative Relativistic Fluid Dynamics}},  {\em
  Phys. Rev.} {\bf C83} (2011) 024906,
  [\href{http://xxx.lanl.gov/abs/1010.4359}{{\tt arXiv:1010.4359}}].

\end{thebibliography}\endgroup
%%%%%%%%%%%%%%%%%%%%%%%%%%%%%%%%%%%%%%%%%%%%%%%%%%%%%%%%%%%%%%%%%%%%%%%%%%
%%%%%%%%%%%%%%%%%%%%%%%%%%%%%%%%%%%%%%%%%%%%%%%%%%%%%%%%%%%%%%%%%%%%%%%%%%
\end{document}